\documentclass[10pt]{article}

\usepackage{amsmath}
\usepackage{amssymb}

\usepackage{graphicx}

\usepackage{cite}

\usepackage{color}

\usepackage[labelfont=bf,labelsep=period,justification=raggedright]{caption}

\makeatletter
\renewcommand{\@biblabel}[1]{\quad#1.}
\makeatother

\date{}

\begin{document}

\begin{flushleft}
{\Large \textbf{Restoration Ecology: Two-Sex Dynamics and Cost Minimization}}
\\
F. Moln\'{a}r Jr.$^{1,\ast}$,
C. Caragine$^{1}$,
T. Caraco$^{2}$,
G. Korniss$^{1}$
\\
\bf{1} Department of Physics, Applied Physics, and Astronomy, Rensselaer Polytechnic Institute, 110 8$^{th}$ Street, Troy, NY, 12180-3590, USA
\\
\bf{2} Department of Biological Sciences, University at Albany, Albany, NY, 12222, USA
\\
$\ast$ E-mail: molnaf@rpi.edu
\end{flushleft}

\section*{Abstract}
We model a spatially detailed, two-sex population dynamics, to study
the cost of ecological restoration.  We assume that cost is
proportional to the number of individuals introduced into a large
habitat.  We treat dispersal as homogeneous diffusion. The local
population dynamics depends on sex ratio at birth, and allows
mortality rates to differ between sexes.  Furthermore, local density
dependence induces a strong Allee effect, implying that the initial
population must be sufficiently large to avert rapid extinction.  We
address three different initial spatial distributions for the
introduced individuals; for each we minimize the associated cost,
constrained by the requirement that the species must be restored
throughout the habitat.  First, we consider spatially inhomogeneous,
unstable stationary solutions of the model's equations as plausible
candidates for small restoration cost.  Second, we use numerical
simulations to find the smallest cluster size, enclosing a spatially
homogeneous population density, that minimizes the cost of assured
restoration.  Finally, by employing simulated annealing, we minimize
restoration cost among all possible initial spatial distributions of
females and males. For biased sex ratios, or for a significant
between-sex difference in mortality, we find that sex-specific
spatial distributions minimize the cost.  But as long as the sex
ratio maximizes the local equilibrium density for given mortality
rates, a common homogeneous distribution for both sexes that spans a
critical distance yields a similarly low cost.


\section*{Introduction}
Ecological restoration aims to replenish an ecosystem's
biodiversity, often responding to human-induced losses of indigenous
species \cite{Lindenmayer_2002,Hall_2012}. When ecosystem managers
reintroduce a species to its former habitat, the restoration
effort's success is ordinarily defined by combined ecological and
economic criteria \cite{Holl_2000}.  Similarly, optimizing
biological-control programs may integrate impact on the target
species with costs of deploying the control agent \cite{Shea_2000}.

Consider an example where restoration failed. Historical records
indicate that Canada lynx (\textit{Lynx canadensis}) were found in
New York State (NYS), but were seen only rarely during most of the
20th century \cite{DEC_2012}.  Between 1989 and 1992, no fewer than
80 lynx were captured in Canada and released in the Adirondack
Mountains of NYS.  Each animal carried a radio-collar, so that
survival and dispersal could be monitored.  The lynx rapidly
dispersed; mortality during dispersal was high.  Lynx population
density grew too low for successful reproduction, and the species is
now considered extirpated in NYS \cite{DEC_2012}.

Generalizing the example, we envision restoration of a single
species whose population dynamics depends on the density of each
sex. Before evaluating costs, we must identify those spatial
distributions of the initial population that assure successful
restoration. Suppose that we initiate restoration with a single
spatial cluster, within which individuals are distributed at a
uniform density.  Then we must find the ``critical cluster'' size,
the minimal area the species must occupy to sustain positive
population growth.  Analysis of the critical-cluster criterion has
advanced understanding of spatial systems in both physics
\cite{Rikvold_PRE1994,Ramos_PRB1999,Machado_PRE2005} and ecology
\cite{Gandhi_JTB1999,KC_JTB2005,OBYKAC_TPB2006,ACK_EER2007}.
However, if ecosystem managers can vary initial densities according
to location, non-uniform spatial distributions might reduce
restoration cost.  Given multiple initial population distributions
assuring sustained population increase, the most preferred option
should \emph{minimize} cost.  Our study investigates how the minimum
cost of successful restoration depends directly on spatial pattern,
and how the optimal pattern depends on sex ratio, and on
sex-specific mortality rates.

In this context, we model a species' restoration as a spatially detailed, two-sex reaction-diffusion system.  We optimize the initial densities and spatial distributions of the sexes to minimize the cost of restoring the species to its positive, stable (homogeneous) equilibrium density throughout a habitat. The prototype of such models is the Fisher-Kolmogorov equation \cite{Fisher_1930,Holmes_1994}
\begin{equation}
\partial_{t}u = D \nabla^{2}u + \alpha u (1-u) \;,
\label{eq1}
\end{equation}
which describes the dynamics of single-sex populations with logistic growth and diffusive dispersal. Our model extends the basic reaction-diffusion framework to include sex-structured dynamics
\cite{Tainaka_EPL2006,Schmickl_EM2010,Molnar_PLOS2012}, where an Allee effect \cite{LK_TPB1993,Keitt_2001} generates an unstable fixed point between extinction and the habitat's carrying capacity.

In many natural and managed populations, \emph{per-capitum} growth is reduced as density becomes small; this is termed an Allee effect \cite{BB_JTB2002,Aviles_2002,BerecTREE_2006}. Different behavioral, ecological and genetic mechanisms can induce an Allee effect \cite{Courchamp_1999}. Low population density may diminish individual reproduction by reducing mate encounters, making prey capture more difficult, or by leaving individuals more susceptible to their own predators \cite{Caraco_1995,Stephens_1999}. Allee effects will be amplified if dispersal into unoccupied habitat reduces local population density; a sufficiently high dispersal rate can generate negative population growth, thwarting restoration \cite{LK_TPB1993}.

Generically, the cost of restoration can be can defined as:
\begin{equation}
C = c^{*}\int_{\Omega} u(r, t=0) dr,
\label{eq2}
\end{equation}
where $\Omega$ represents the extent of the habitat, and $u(r,t)$ is the population density at location $r$.  Without loss of generality, we can consider a constant \emph{per-capitum} cost ($c^{*}=1$). Mathematically, the restoration cost, which we seek to minimize, is a functional of the initial population's spatial distribution.  The constraint requiring population persistence cannot be expressed analytically, since that would require solving the model's partial differential equations exactly. Therefore, we develop a population dynamics with simple processes and easily interpreted parameters, and obtain the minimum cost with analytical and numerical techniques.

We organize the rest of the paper as follows. First, we introduce our sex-structured population dynamics, and outline the analytic and numerical methods we employ in this paper. In the Results section we conduct a systematic study of the restoration cost in multiple stages. Starting with a simplified, single-sex version of our model, we derive an unstable, aperiodic stationary solution for the PDE, resulting in a single-sex cost. We then continue with the sex-structured model and analyze the way restoration cost depends on model parameters, given a simple, constrained (i.e. homogeneous) initial distribution of individuals. Thereafter, we relax the constraints in two steps, and study how the cost can be reduced as a result. Finally, in the Discussion we compare the minimum costs found in each stage, and conclude which approach yields the most economical restoration.


\section*{Methods}

\subsection*{General assumptions}

Our model's key parameters include the sex ratio at birth and sex-specific mortality rates.  We assume that females and males disperse independently by homogeneous diffusion, and that males encounter females as a mass-action process, equivalent to random mating \cite{BBB_AmNat2001}. The fraction of matings leading to successful reproduction is proportional to the unoccupied fraction of the environment, $(1 - m - f)$
\cite{Tainaka_EPL2006,Schmickl_EM2010,Molnar_PLOS2012}. That is, the population grows in a self-regulated manner. Hence, we have:
\begin{eqnarray}
\partial_{t} f & = & D_{f} \nabla^{2}f + \theta \left(1-m-f\right) f m - \mu_{\rm f} f \nonumber \\
\partial_{t} m & = & D_{m} \nabla^{2}m + \left(1-\theta\right) \left(1-m-f\right) f m - \mu_{\rm m} m \;,
\label{eq-model}
\end{eqnarray}
where $f(x,t)$ and $m(x,t)$ denote the local densities of females
and males, respectively. Diffusion rates are described by
coefficient $D_f$ for females and $D_m$ for males. Three parameters
characterize the local dynamics: $\theta$, $\mu_{m}$, and $\mu_{f}$,
which denote the fraction of individuals born female, and
density-independent mortality rates for males and females,
respectively. Note that while Eqs.~(\ref{eq-model}) do incorporate
spatial effects, they retain a ``mean-field" character (the
statistical physics terminology) in that correlation functions of
the underlying stochastic individual-based model are factorized into
products of densities \cite{McKane_PRE2004,Korniss_EPL1995}. In
principle, by extending the above deterministic reaction-diffusion
equations with appropriate noise terms
\cite{Korniss_PRE1997,Escudero_PRE2004,Pigolotti_TPB2013}, the
resulting stochastic partial differential (or Langevin-type)
equations \cite{Gardiner_1985} can capture the relevant macroscopic
features of the underlying spatial, stochastic individual-based
model \cite{vanKampen_1981,Schmittmann_1995,Hinrichsen_2000}. (Note,
however, that the rigorous derivation of such stochastic partial
differential equations can be rather challenging
\cite{Doi_JPA1976,Tauber_JPA2005,Tauber_JPA2012}.)

Our model [Eq.~(\ref{eq-model})], with spatial detail and diffusion
removed, is identical to that studied by Tainaka et al.
\cite{Tainaka_EPL2006}. In our earlier work \cite{Molnar_PLOS2012},
we briefly analyzed the above two-sex dynamics with diffusive
dispersal, and found the critical radius (hence, critical cluster
size) for an initial population's successful invasion of a
two-dimensional habitat. While our reaction-diffusion model includes
numerous simplifications for detailed application to particular
species, nevertheless it exhibits the essential ecological
characteristics of more complex two-sex models. Hence, implications
of our results for restoration will likely hold across a wide range
of specific models.

We can transform Eq.~(\ref{eq-model}) to interpret it as a
single-sex model by making the equations symmetric. To do so, we must let $\theta=0.5$, restrict $\mu_{f}=\mu_{m}=:\mu$,
$D_{f}=D_{m}=:D$ and use the same initial-density distributions for both males and females. In this way, the two densities behave identically over time: $f(x,t) = m(x,t) = u(x,t)$,
described by the following equation:
\begin{equation}
\partial_{t} u = D \nabla^{2}u + \frac{1}{2}u^{2}(1-2u) - \mu u.
\label{eq-simplemodel}
\end{equation}
This transformation bridges the single-sex and two-sex models; using the constants in Eq.~(\ref{eq-simplemodel}), we can directly compare results between models without rescaling parameters. Note that the (cubic) local dynamics also retains an Allee effect. 

\subsection*{Analytic and numerical methods}
Our first approach to cost minimization selects suitable unstable stationary solutions of the PDE model as initial population distributions, since we can derive them analytically for the single-sex model.  These are special ``critical'' solutions, which transform to stable equilibrium solutions (either persistence or extinction) depending on a small perturbation. The cost associated with each stationary solution is found by numerical integration of the density profile (i.e., the area under the curve). We obtain the stationary solutions for the single-sex model by setting the left-hand side of Eq.~(\ref{eq-simplemodel}) to zero, and deriving a relationship between the stationary solution's local value and its derivative.

For the sex-structured model, we cannot derive stationary solutions analytically. Instead, we base our study on the model's critical limits of cluster size and density, which are directly related to the minimum cost of successful restoration. For a homogeneous initial spatial distribution with a specific population density, there exists a critical cluster size, the smallest spatial extent such that the given density achieves sustained positive growth. Symmetrically, for a given cluster size, there is a critical initial density, the lowest density assuring sustained population growth. In both cases, the critical limit also corresponds to the minimum cost, since the cost is proportional to both cluster size and density. The exact, parameter-dependent values of these critical limits cannot be derived analytically. Instead, we use binary search across a range of possible values, accomplished by testing each value for successful restoration \emph{vs} extinction.

We discriminate restoration from extinction by numerically integrating the model until it has converged to a global equilibrium.  We use second-order finite difference discretization for spatial derivatives and explicit Euler method for integration over time \cite{method_of_lines}, with a sufficiently small time-step. Integration stopped when all time-derivatives at all spatial coordinates were less than $10^{-6}$. Finally, the cost is obtained by multiplying the initial density and the initial cluster size, and summed for males and females.

Next, we extend the uniform-density approach by allowing the initial cluster size to differ between males and females. In this case, we first calculate the critical cluster size for males at a fixed cluster size for females using binary search, resulting in a cost with respect to the given female cluster size. Then, we use gradient descent on this cost function to minimize it with respect to female cluster size. Note, that during the gradient descent we always change the female cluster size by one unit of spatial grid resolution, and we keep moving toward the negative gradient even if the local derivative is zero, because a small slope discretized with any grid can result in zero local gradients before reaching the actual minimum point. Note also, that we strongly rely on the convexity of the cost function with respect to male and female cluster sizes.

Finally, we relax all constraints on the shape of the initial population distribution; we optimize the spatially discretized shape for lowest cost under the constraint of successful restoration. Discretization is essential because it allows us to express the cost as an $n$-dimensional function, instead of a functional, where $n$ is the size of the spatial grid. We use the same grid for shape discretization and numerical PDE integration, for practical reasons. Given the discretization, we use \textit{simulated annealing}\cite{sim-anneal} for optimizing the shape function. This is essentially a Monte Carlo simulation, where random changes of the initial spatial distribution (the shape function) are accepted or rejected according to a specified acceptance probability function, such that the visited cost states have a Boltzman-distribution characterized by a temperature-like parameter. As this parameter is lowered, the expected value of the cost is also lowered, eventually leading to the globally optimal, minimum cost state. For the specific steps of simulated annealing, see Supporting Information (Section S5).  In order to determine whether the constraint of successful restoration is satisfied we must numerically integrate the model at every Monte Carlo step. To reduce computational time, we accelerated the PDE integration with GPGPU computation using CUDA \cite{ref-cuda,cuda-pde1}, which locates the global equilibrium of the PDE in a fraction of a second, giving a total time for the simulated annealing in the order of a few hours.

\section*{Results}

The prerequisite for analyzing the cost of restoration is to ensure the local stability of successful restoration, i.e., a positive stable fixed point of local dynamics. The necessary stability condition for the two-sex model [Eq.~(\ref{eq-model})] is given in our previous work \cite{Molnar_PLOS2012}:
\begin{equation}
1 - 4\left( \frac{\mu_{f}}{\theta} + \frac{\mu_{m}}{1 - \theta} \right) > 0
\label{eq-stability}
\end{equation}
Similarily, we can find the necessary stability condition for the single-sex model [Eq.~(\ref{eq-simplemodel})]:
\begin{equation}
\mu < \frac{1}{16}.
\label{eq-stability-simple}
\end{equation}
These conditions provide us guidelines for selecting proper model parameters when evaluating costs in the following sections. Justification for Eqs.~(\ref{eq-stability}) and (\ref{eq-stability-simple}), and formulas for the fixed points are presented in the Supporting Information (Sections S1 and S2). 

\subsection*{Unstable stationary solutions}
An unstable stationary solution of the PDE seems a good candidate for the initial spatial distribution. It is a critical solution, in the sense that given a small perturbation, the unstable stationary solution transforms to a stable, spatially homogeneous solution. Positive perturbations result in positive homogeneous population densities (successful restoration), and negative perturbations result in zero densities (extinction).

\begin{figure}[t!]
\begin{center}
\includegraphics[width=\textwidth]{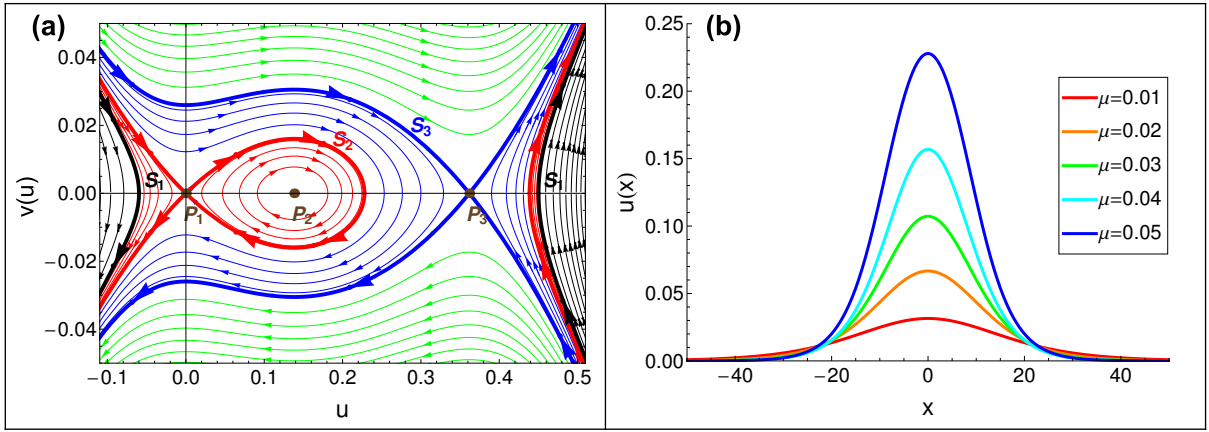}
\caption{(a) Phase plot of stationary solutions in the single-sex model, described by Eq.~\ref{eq-vusol}. $D=1.0$, $\mu=0.05$. The dots indicate fixed points, the thick lines indicate separatrices. Different curves correspond to different $E$ parameters, however, values of $E$ were not chosen uniformly, for aesthetic reasons. (b) The stationary solutions found by integrating along the $S_{2}$ separatrix, for multiple mortality rate parameters; $D=1.0$; $x$ is distance form the habitat's center.}
\label{fig-uv}
\end{center}
\end{figure}

Partial differential equations, such as Eqs.~(\ref{eq-model}) and
(\ref{eq-simplemodel}) have infinitely many unstable stationary solutions. We cannot find these solutions directly, but we can derive analytical formulas for the relationship between the density and its spatial derivative. For the single-sex model
[Eq.~(\ref{eq-simplemodel})] we have the following definitions:
\begin{eqnarray}
\frac{\partial u}{\partial x} & = & v  \\
\frac{\partial v}{\partial x} & = & \frac{-1}{D} \left( \frac{1}{2}u^{2}(1-2u) - \mu u \right). \;
\label{eq-defv}
\end{eqnarray}
Using $v$, we change variables to write a first-order differential equation:
\begin{equation}
v(u) \frac{\partial v}{\partial u} = \frac{-1}{D} \left( \frac{1}{2}u^{2}(1-2u) - \mu u \right).
\label{eq-vu}
\end{equation}
By separating variables, we obtain the following analytical solution:
\begin{equation}
v(u) = \pm \sqrt{ \frac{3u^{4} - 2u^{3} + 6 \mu u^{2} + 12 D E}{6D} },
\label{eq-vusol}
\end{equation}
where $E$ is a free parameter. The phase diagram for this equation is depicted in Fig.~\ref{fig-uv}(a), and its contents are summarized as follows.

The fixed points $P_i$ correspond to \textit{homogeneous} stationary solutions. Naturally, these are also fixed points of the original equations [Eq.~(\ref{eq-simplemodel})], but here they are only special cases of stationary solutions that do not vary spatially.  Hence $v(u)=0$.  $P_{1}$ and $P_{3}$ are saddle points (stable equilibrium nodes of the local dynamics) corresponding to extinction ($u=0$) and persistence ($u>0$), respectively. $P_{2}$ is a center (unstable fixed point of the local dynamics) corresponding to the unstable equilibrium due to the local dynamics' strong Allee effect.

Curves in Fig. \ref{fig-uv}(a) correspond to \textit{inhomogeneous}
stationary solutions that can be classified by the value of the free parameter $E$ in Eq.~(\ref{eq-vusol}). Separatrices $S_1$, $S_2$, and $S_{3}$ correspond to the following values (with the same subscripts, see Supporting information Section S3 for details):
\begin{eqnarray}
E_{1} & = & \frac{1}{48D} (u^{*}_{2} - 2 \mu) (u^{*}_{2} - 6 \mu)\\
E_{2} & = & 0\\
E_{3} & = & \frac{1}{24D} (u^{*}_{3})^{2} (u^{*}_{3} - 6 \mu),
\label{eq-Evalues}
\end{eqnarray}
where $u^{*}_{1}$, $u^{*}_{2}$, and $u^{*}_{3}$ are the equilibrium densities of local dynamics, i.e., the values of $u$ corresponding to $P_1$, $P_2$, and $P_3$, respectively. The closed elliptical curves around $P_{2}$ represent periodic stationary solutions, and they are the only ones of interest, because all other curves extend to infinitely large negative or positive densities, neither having biological meaning.

Spatially periodic stationary solutions may offer candidate initial population distributions. In principle, if minimum densities within each period were close to zero, then we could select a segment of the solution, one period in length between two density minima, and apply it as an initial spatial distribution. However, in our case, as the minimum value of $u$ goes to zero, the period of the solutions goes to infinity, and the curves converge to the $S_2$ separatrix, which corresponds to an aperiodic stationary solution. The exact aperiodic shape of $u(x)$ can be found by numerical integration along $S_2$, depicted in Fig.~\ref{fig-uv}(b). Since $u(x)$ converges to zero rapidly, we can use it as an initial population distribution by taking its central segment above an arbitrary small (biologically meaningful) density threshold. Because of the fast convergence to zero, the length of the segment will be finite. In our study, we use $10^{-5}$ for the density threshold. Then, for every $D$ and $\mu$ combination, we have an exact shape, and an exact cost value defined as twice the area under the curve (counting both males and females), denoted by $C_{\textrm{stat}}$. We will compare these costs with those found by the other two methods described in the following subsections.

As an interesting observation, we note that contrary to intuition, the period length of the stationary solution does not converge to zero as the solution curves approach $P_{2}$, see Supporting Information (Section S4).

For the sex-structured model [Eq.~\ref{eq-model}] we cannot derive unstable stationary solutions analytically. Instead, we study the scaling of restoration cost through a numerical analysis of the critical cluster size.


\subsection*{Critical cluster size and minimum cost}
Criticality of an initial cluster's size occurs when density reduction due to dispersal exactly balances the net effect of local natality and mortality; Supporting Video S1 clearly shows this form of criticality in one dimension. In two dimensions the expanding population front's speed is reduced in proportion to the curvature of the cluster; therefore, it affects the critical cluster size. To avoid confusing effects of curvature with other parameters' impact on the critical cluster size and, hence, the minimum cost, we restrict our study to one-dimensional space.

We begin our analysis of the critical cluster size by assuming a uniform spatial distribution. This is the most obvious choice, for its mathematical simplicity, and its plausible application (\emph{e.g}., an animal population surrounded by a fence before release can be modeled with a uniform spatial distribution). Therefore, we have four parameters describing the distribution: $f_{0}$, $m_{0}$, $l_{f}$, $l_{m}$, which represent female density, male density, length of space occupied by females, and length of space occupied by males, respectively, at the initiation of restoration. We refer to this initial setup as ``rectangular'', for its shape on a density \emph{vs} location plot, and we shall refer to $l_{f}$ and $l_{m}$ as the cluster sizes of the initial population. The cost of the initial state is defined simply as:
\begin{equation}
C_{\text{rect}} = f_{0} \times l_{f} + m_{0} \times l_{m},
\label{eq_cost1}
\end{equation}
and it is minimized by using the critical cluster sizes $l_{m}^{*}$ and $l_{m}^{*}$, for males and females, respectively. Note, that $l_{m}^{*}$ and $l_{m}^{*}$ are themselves dependent on initial densities $f_{0}$ and $m_{0}$, as well as model parameters $D_{f}$, $D_{m}$, $\mu_{f}$, $\mu_{m}$, and $\theta$.  Therefore, we systematically study dependence of critical clusters on all parameters.

As an initial step, we analyze the dependence of critical cluster size on the diffusion coefficients. We anticipate that the critical cluster size, \emph{i.e}., critical length (in one-dimensional space) is proportional to the square root of the diffusion coefficient. Similar scaling has been observed in two-dimensions when a population with a strong Allee effect disperses by diffusion \cite{LK_TPB1993,Molnar_PLOS2012}. Our aim here is to show the same behavior in one-dimensional space, and to ask whether it holds when male and female diffusion coefficients and cluster sizes differ.
\begin{figure}[t!]
\begin{center}
\includegraphics[width=\textwidth]{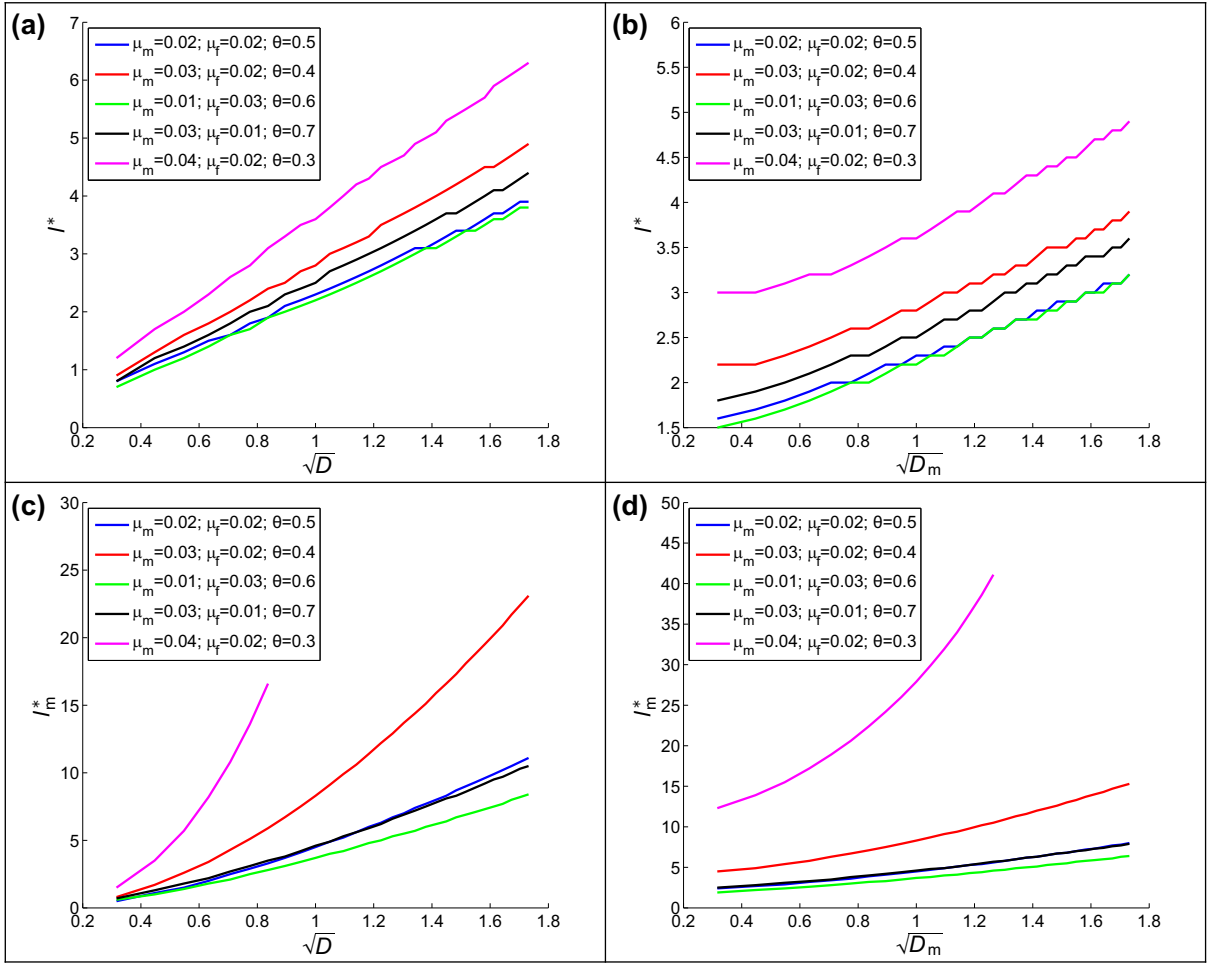}
\caption{Scaling of critical cluster sizes \emph{vs} diffusion coefficients, at various parameter values. (a) $l_{f}^{*} = l_{m}^{*} = l^{*}$, $D_{f}=D_{m}=D$; (b) $l_{f}^{*} = l_{m}^{*} = l^{*}$, $D_{f}=1.0$; (c) $l_{f}^{*}=1.0$, $D_{f}=D_{m}=D$; (d) $l_{f}^{*}=1.0$, $D_{f}=1.0$. In every case, initial (spatially homogeneous) densities are fixed: $f_{0} = m_{0} = 0.45$.}
\label{fig-rc-diff}
\end{center}
\end{figure}

Figure \ref{fig-rc-diff}(a,b) shows that as long as we employ the same cluster size for males and females, the critical cluster size has the expected scaling behavior $l^{*}\sim\sqrt{D}$ \cite{LK_TPB1993} with respect to both male and female diffusion coefficients, even if one sex has a fixed diffusion coefficient.  Note that it is sufficient to study the dependence on the diffusion of one sex (here, males) while the other is fixed (here, females), because of the symmetric construction of the model. Further,
Fig.~\ref{fig-rc-diff}(c,d) indicates that fixing the cluster size of one sex while measuring the critical cluster size of the other sex with respect to diffusion coefficients yields non-trivial scaling. However, we observe that a higher dispersal rate results in a larger critical cluster size, and, hence, a larger restoration cost.

Continuing our analysis of critical cluster sizes, we now assume that the critical cluster sizes are equal for both sexes ($l_{f}^{*}=l_{m}^{*}=:l^{*}$); we relax this constraint later. Even with the equal cluster-size constraint, initial densities for males and females within that cluster may, in principle, differ.  In practice, a density difference could be implemented for most diecious species. To find the best choice of initial density values (with respect to minimizing cost), we aim to relate them to model parameters, taken as given for the focal population.

\begin{figure}[t!]
\begin{center}
\includegraphics[width=\textwidth]{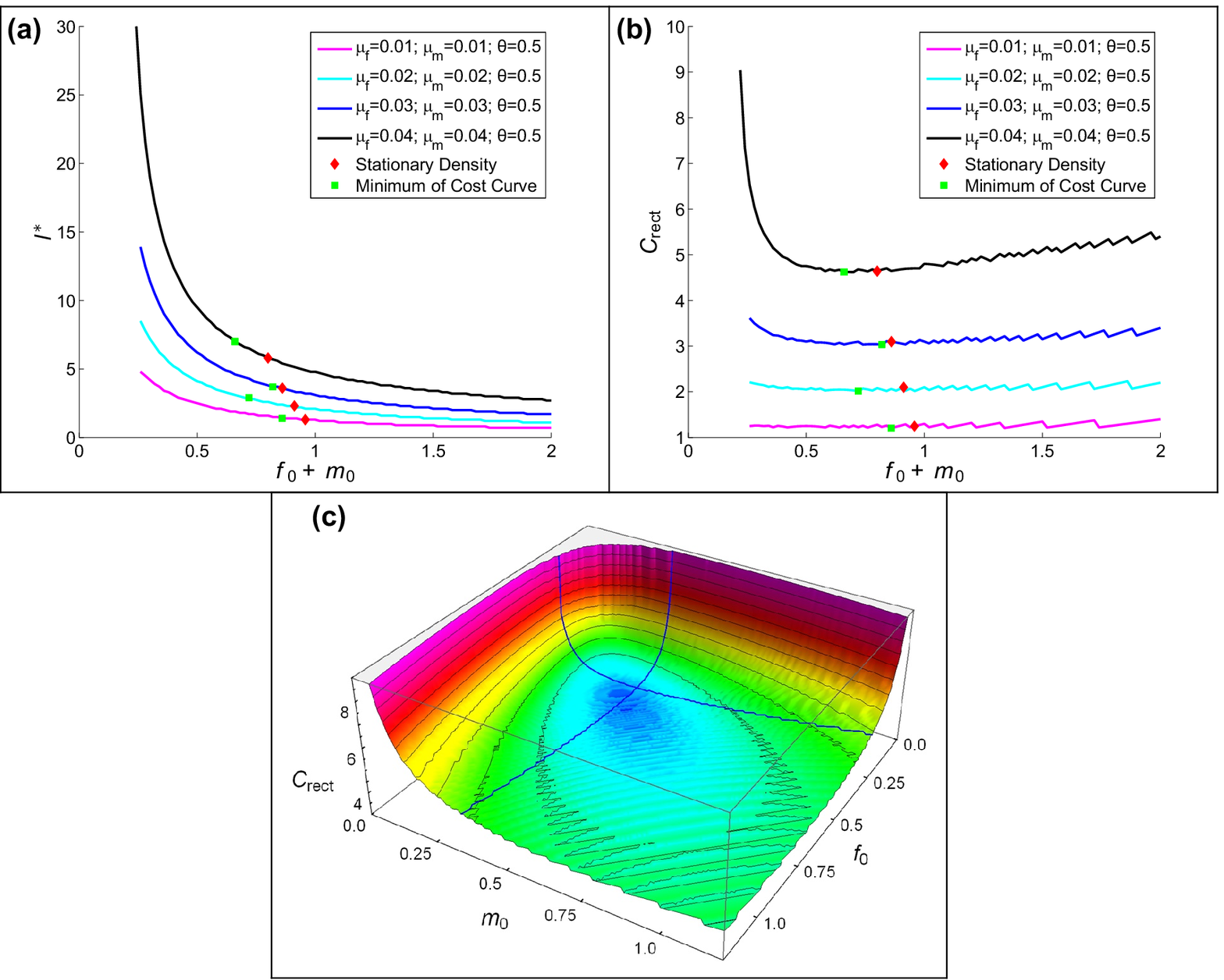}
\caption{Scaling of (a) critical cluster length at introduction, and (b) cost, with respect to initial population density. The total density shown on the $x$ axis is divided equally between males and females. The point markers on the curves show the stable stationary densities of local dynamics. (c) shows the cost landscape with respect to male and female densities; the blue cross marks the stable stationary density values; $\mu_{f}=0.04$, $\mu_{m}=0.03$, $\theta=0.5$. For all figures, $D=1.0$. }
\label{fig-vs-dens}
\end{center}
\end{figure}

Naturally, the initial densities must exceed the Allee threshold (the unstable fixed-point densities); otherwise, the population can never achieve positive growth. Figure \ref{fig-vs-dens}(a) shows that as the initial density is lowered, the critical cluster size increases, and goes to infinity as we approach the Allee threshold.

Scaling of the cost, however, is non-trivial. We can always find the minimum cost at a density somewhere in the vicinity of the positive stable fixed point of the system, but always slightly below it. We understand this by considering the dynamics just after initial introduction. If the population starts from its stationary density (the stable, positive fixed point), then the local densities can only decrease, due to diffusive dispersal. However, if the initial density is lowered slightly, then the population has a chance to grow locally (in particular, at the center of the cluster) before the effects of diffusion reach it, while the eventual spread through the habitat remains the same. In essence, the cost is slightly lowered by handing over some of the spreading effort to growth dynamics. However, as Fig.~\ref{fig-vs-dens}(b) and Fig.~\ref{fig-vs-dens}(c) show, this advantage in cost-reduction is very small. We can conclude that using the stationary densities as initial densities results in a sufficiently low cost. Also, it provides a good choice of initial density based on model parameters, because the stationary densities depend on local dynamics, which in turn depend on model parameters. The formulas for the stationary densities are included in the Supporting Information (Sections S1 and S2).

We expand on the cost-minimizing property of stationary densities; we use them as initial densities throughout the rest of our study. We continue with the analysis of the critical cluster size's dependence on model parameters. In particular, we focus on the value of sex ratio at birth ($\theta$), because both population stability and equilibrium density depend on this single parameter. For population stability and persistence, there is a range of permissible sex ratios, determined by the mortality rates (see Supporting Information Section S2).  Within this range, we define the optimal sex ratio $\theta^{*}$ as the value maximizing the equilibrium population density \cite{Tainaka_EPL2006,Molnar_PLOS2012}:
\begin{equation}
\theta^{*} = \frac{1}{1+\sqrt{\mu_{f}/\mu_{m}}}
\label{eq-optimal-theta}
\end{equation}
Note, that when the mortality rates are equal, the optimal sex ratio is $0.5$, which is the parametrization, by definition, in the symmetric single-sex model.

\begin{figure}[t!]
\begin{center}
\includegraphics[width=\textwidth]{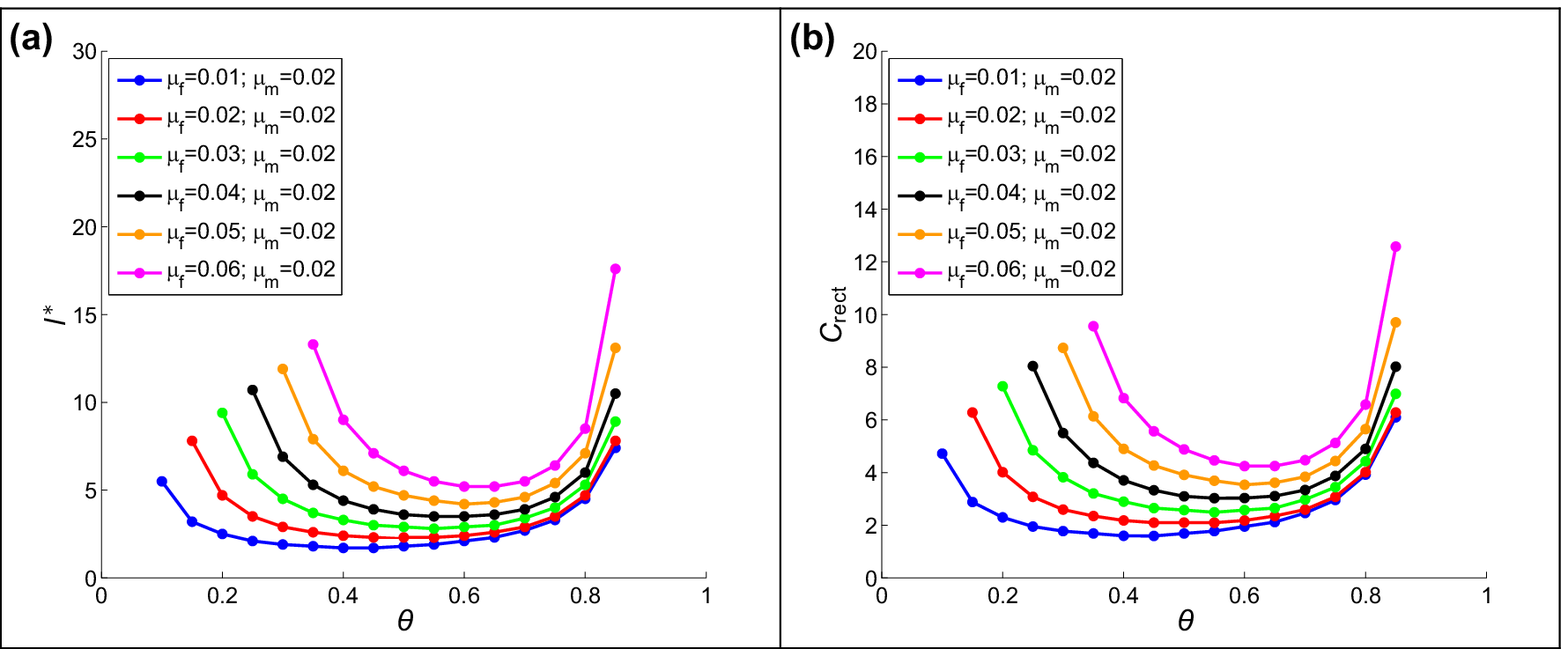}
\caption{ Scaling of (a) critical cluster size and (b) cost, with respect to sex ratio, at different mortality rate combinations; $D=1.0$. Rectangular initial populations were used with stationary population densities.}
\label{fig-vs-theta}
\end{center}
\end{figure}

We find that the smallest cluster size assuring restoration corresponds closely to the optimal sex ratio, and that any small deviation from the optimal value causes a small increase in the critical cluster size [Fig.~\ref{fig-vs-theta}(a)]. However, since the equilibrium population densities (serving as initial densities) decrease at suboptimal sex ratios (by definition; see Fig.~S1 in Supporting Information), the combined effect on the cost is non-trivial. As we see on Fig.~\ref{fig-vs-theta}(b), restoration cost is minimized at approximately the same sex ratios minimizing the critical cluster size, indicating that the cluster size is more sensitive to biased sex ratios than to equilibrium densities. We also find that strongly biased sex ratios approaching the boundary of the stability range cause both cluster size and restoration cost to diverge.

To complete the relationship between the sex ratio minimizing restoration cost ($\widehat{\theta}$) and the sex ratio that locally maximizes total equilibrium population density ($\theta^{*}$), we compare the two quantities numerically. For the latter we have an analytical expression
[Eq.~(\ref{eq-optimal-theta})].  But our cost-minimizing sex ratios have only limited precision, since for each value of $\theta$ we employed binary search to determine the critical cluster size, which, in turn, determines the cost. Therefore we define a computational error bound on $\widehat{\theta}$ as the range of $\theta$ values that give critical cluster sizes within the error range of the minimum point's cluster size found by binary search.

\begin{figure}[t!]
\begin{center}
\includegraphics[width=\textwidth]{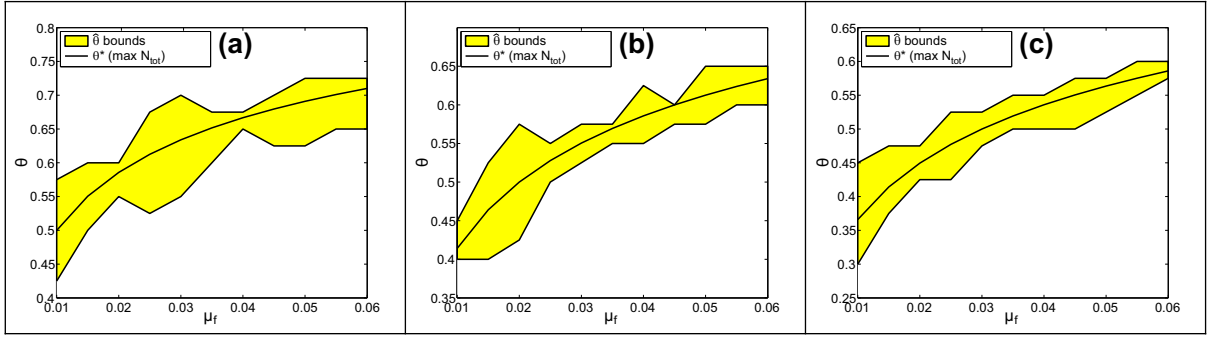}
\caption{Comparision of density-maximizing sex ratios $\theta^{*}$ [Eq.~(\ref{eq-optimal-theta})] and numerical bounds of cost-minimizing sex ratios $\widehat{\theta}$, using homogeneous population distributions with stationary initial densities, $D=1.0$, (a) $\mu_{m}=0.01$, (b) $\mu_{m}=0.02$, (c) $\mu_{m}=0.03$.}
\label{fig-theta-vs-mu}
\end{center}
\end{figure}

Figure \ref{fig-theta-vs-mu} offers comparison of the
density-maximizing and cost-minimizing sex ratios. We conclude that
the sex ratio maximizing equilibrium density is identical to the sex ratio minimizing restoration cost, up to computational error.


To this point, our results reflect the assumption that individuals of each sex are introduced across the same extent of habitat. We now relax this constraint; that is, we permit $l_{f}^{*} \neq l_{m}^{*}$, and ask whether the cost of restoration can be reduced by introducing individuals into sex-specific lengths of habitat. We denote the ratio of costs obtained by unequal and equal cluster sizes as:
\begin{equation}
c_{\textrm{rel}} = \frac{C_{\textrm{rect}}[l_{f}^{*} \neq l_{m}^{*}]}{C_{\textrm{rect}}[l_{f}^{*} = l_{m}^{*}]}
\label{eq-crel}
\end{equation}

\begin{figure}[t!]
\begin{center}
\includegraphics[width=\textwidth]{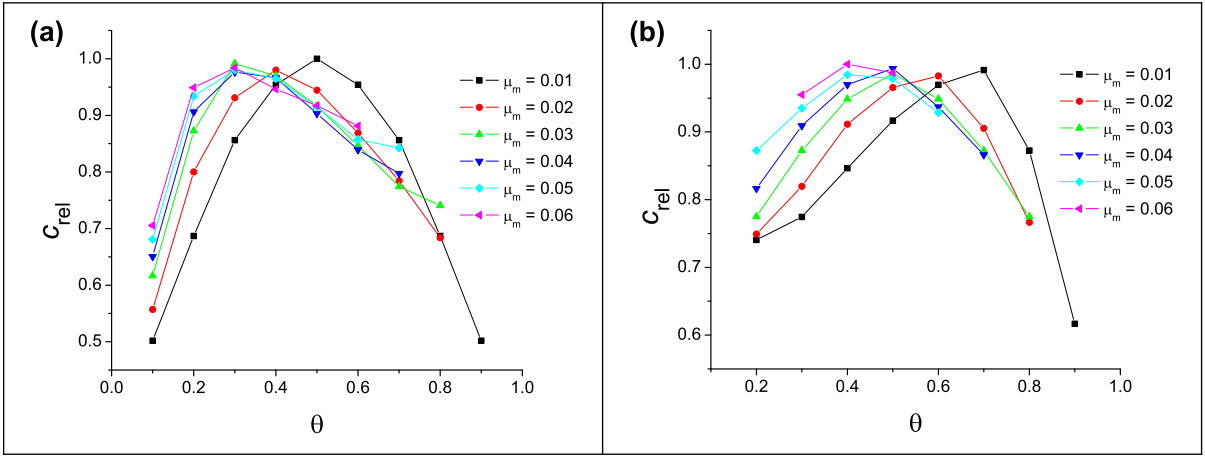}
\caption{Minimum cost found by allowing different initial cluster sizes for males and females, relative to the case where cluster lengths are equal. Common parameter: $D=1.0$. Individual parameters: (a) $\mu_{f}=0.01$, (b) $\mu_{f}=0.03$.}
\label{fig-relcost}
\end{center}
\end{figure}

Figuse~\ref{fig-relcost} presents our results. We find that optimality of the sex ratio plays a central role in reducing the cost of restoration. In particular, when the sex ratio equals the optimal (density-maximizing) value determined by mortality rates, the minimum cost is achieved with equally sized clusters.  No further cost reduction can be achieved by allowing different cluster sizes for males and females. However, the relative advantage of sex-specific cluster size increases as the sex ratio deviates from the optimal value. If we consider that the absolute cost value diverges for strongly biased sex ratios [see Fig.~\ref{fig-vs-theta}(b)] we conclude that in such cases the savings achieved by adjusting the initial cluster sizes of each sex could be substantial.

\subsection*{Simulated annealing}

We now relax all constraints on the initial cluster's spatial distribution, and use simulated annealing to optimize sex-specific distribution shapes with respect to cost. At this stage we assume only that the cost-minimizing distributions have a finite support, and we carry out the minimization accordingly. However, the support of the function is allowed to grow or shrink by random shape changes during simulated annealing; see Supporting Information (Section S5) for details.

\begin{figure}[t!]
\begin{center}
\includegraphics[width=\textwidth]{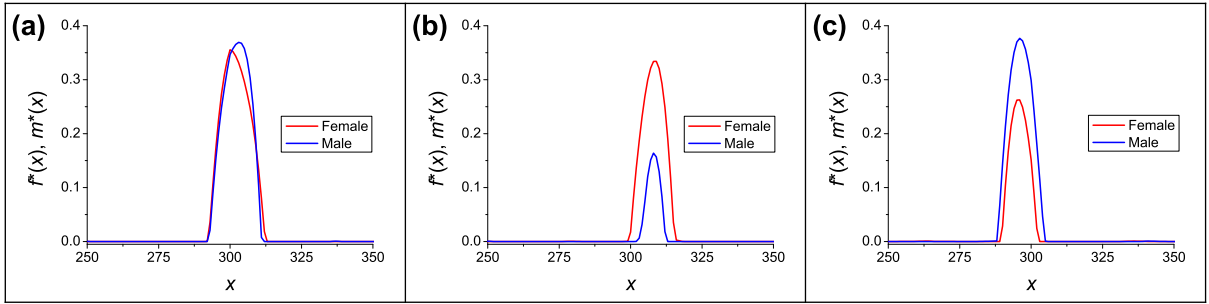}
\caption{Shapes of initial density distributions that minimize cost, found by simulated annealing. The $x$ axis shows discretization grid coordinates; $dx=0.1$ length unit per grid point. Parameters: (a) $\theta = 0.5$, $\mu_{f} = 0.04$, $\mu_{m} = 0.04$, (b) $\theta = 0.1$, $\mu_{f} = 0.01$, $\mu_{m} = 0.02$, (c) $\theta = 0.7$, $\mu_{f} = 0.04$, $\mu_{m} = 0.02$.}
\label{fig-mincost-shape}
\end{center}
\end{figure}

By analyzing a series of minimum cost distributions obtained with simulated annealing, we observe the following. First, the distributions indeed have a finite support. Although we initialize them as such, the width of the optimal population distribution tends to become smaller, rather than larger, during simulated annealing. This effect during the minimization procedure is shown by Supporting Video S2; typical final, optimized shapes are shown in Fig.~\ref{fig-mincost-shape}. It is also remarkable that the edges of the distributions go to zero very sharply; this property develops without any influence inherent to the procedure.

Generally, the final result is an ``arch''-shaped distribution, with similar dimensions for females and males. Note that as the sex ratio diverges from its optimal value, we observe a change in the sizes of the two initial population distributions and in the height of the peaks.  These changes in spatial distributions occur roughly in proportion to the system's positive stationary densities [Fig.~\ref{fig-mincost-shape}(b,c)]. Interestingly, the height of the peaks always falls between the stationary densities and the Allee threshold. Note that this shape provides the maximal rate of population growth possible during the first moments of the simulation, hence it combats the diffusion-amplified Allee effect most efficiently. Costs corresponding to the optimized distributions are denoted $C_{\textrm{SA}}$ when we compare results with other methods.

\section*{Discussion}
We examined three approaches to minimizing the cost of a species' restoration; the approaches differ in both ecological premises and mathematical methods. We considered the aperiodic, spatially inhomogeneous solution to the single-sex dynamics [Fig.~\ref{fig-uv}(b)], critical cluster sizes with spatially homogeneous density [Fig.~\ref{fig-vs-theta}(a)], and simulated annealing of sex-specific initial distributions [Fig.~\ref{fig-mincost-shape}].

\begin{figure}[t!]
\includegraphics[width=3.27in]{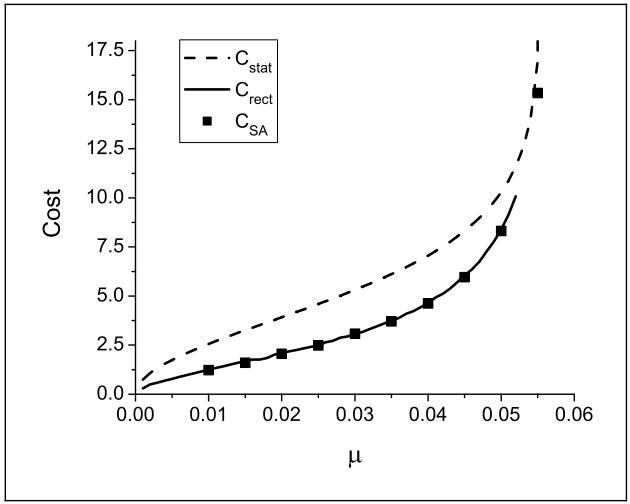}
\caption{Comparison of minimum cost values in the symmetric model, found by integrating the aperiodic stationary density, by using rectangular shape with stable stationary densities, and with simulated annealing.}
\label{fig-mincost-aperiodic}
\end{figure}

Summary comparisons of the minimum restoration cost achieved by the different methods for the symmetric single sex model appear in Fig.~\ref{fig-mincost-aperiodic}. The aperiodic stationary solution to the dynamics gives significantly larger cost than the other approaches. Considering the shape of these distributions, they would likely prove difficult to implement in application. Restricting attention to the other two approaches, it is remarkable that simple homogeneous distributions and the results of simulated annealing yield essentially identical costs.

\begin{figure}[t!]
\begin{center}
\includegraphics[width=\textwidth]{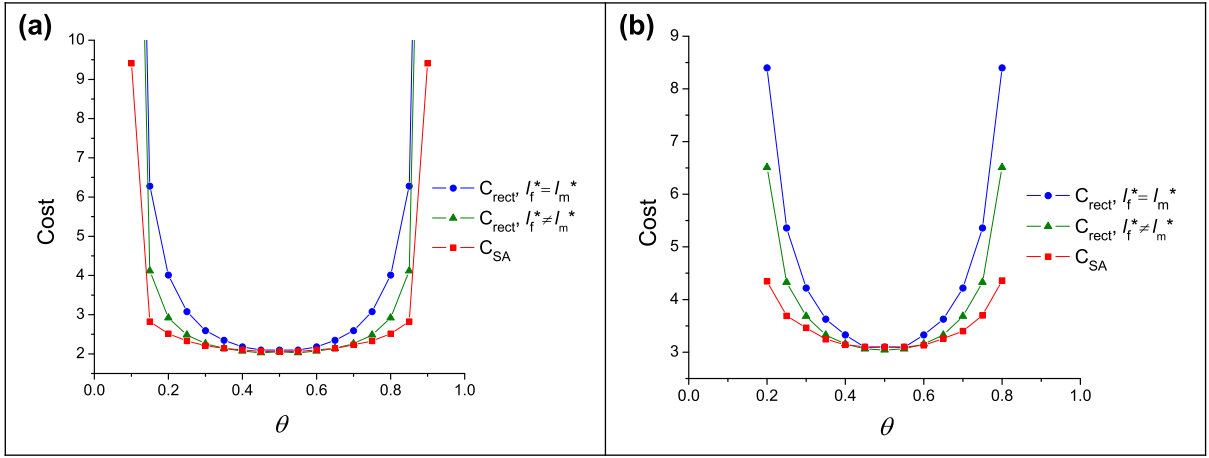}
\caption{Comparision of the cost of successful restoration found by different methods, (a) $\mu_{f}=\mu_{m}=0.02$, (b) $\mu_{f}=\mu_{m}=0.03$.}
\label{fig-mincost-compare}
\end{center}
\end{figure}

Figure \ref{fig-mincost-compare} compares minimum costs for each critical-cluster analysis (i.e., a single cluster size and sex-specific cluster sizes) and costs incurred under simulated annealing. The critical cluster methods assume uniform initial population density within cluster bounds; simulated annealing lets initial densities depend on spatial location. The minimum cost varies little among methods as long as the sex ratio at birth does not deviate too much from the optimal value (here, $\theta^{*} = 0.5$). As sex-ratio bias increases, optimal sex-specific initial cluster sizes can lower the minimum cost of restoration. Simulated annealing reduces restoration cost even further, but this advantage becomes significant only at strongly biased (and biologically rare) sex ratios, and implementing such spatial distributions in application could prove difficult, negating any cost advantage. The same qualitative conclusions hold when we fix the sex ratio and increase the difference between the sexes' respective mortality rates, because the mortality bias can also increase the difference between the optimal and any fixed sex ratio.

Our model assumes deterministic dynamics, which does not account for extinction due to demographic stochasticity in populations near an extinction threshold \cite{Dennis_1989,Lande_1998}. This effect can be exaggerated when a population's spatial dispersion leaves dynamically independent clusters near critical size \cite{Caraco_2001,KC_JTB2005}.

We assume diffusive dispersal. Many plants, and some animals, are ``diffusion limited,'' i.e., the probability of long-distance dispersal is much lower than diffusion assumes \cite{Ellner_1998,OMalley_PRE2006,Omalley_2009}. Diffusion limitation implies that the criterion for expected positive growth demands greater propagation, relative to mortality, as dispersal limitation increases \cite{Caraco_1998,OBYKAC_TPB2006}. We also assume that no
explicit interspecific interactions affect the population during restoration. Species occupying the community to be restored may facilitate restoration; for example, trees may attract birds that disperse seeds of other tree species \cite{Robinson_2000}.  Alternatively, resident species may resist the introduced species biotically \cite{Levine_2004,Allstadt_2009}.  Interspecific interactions will often affect the likelihood of restoration success, as well as the cost.  Consequences of these interactions can sometimes be expressed abstractly through the introduced species' positive equilibrium density; in other cases, successful restoration may demand quantification of these interactions.

We assume an Allee effect arises from interaction of self-regulation with a birth rate that depends on the density of each sex.  In the context of restoration, a two-sex dynamics may be essential to predicting spatial-expansion rate if dispersal differs between sexes \cite{Miller_2011}.  We model mating encounters \emph{via} mass-action, which should be reasonable for animals maintaining individual home ranges, or for dioecious plants with random mating.  Alternative ``marriage functions'' \cite{Caswell_1986} apply to certain species, particularly for polygynous or polyandrous mating systems.

We modeled a single species' restoration only.  Habitat restoration may attempt to manage particular multi-species interactions, or may seek to promote growth of many threatened species \cite{Lindenmayer_2002}.  Our cost function ignores feedback of a species' restoration on other biotic processes, or on economic stake-holders who incur post-restoration costs \cite{Buckley_2006}.

Our results suggest some considerations for species restoration. First, if a species disperses rapidly, individuals should be introduced concurrently, rather than serially.  The initial population will increase only if the density exceeds any Allee threshold, and continues to do so as individuals disperse. Intuitively, the number/density of individuals introduced should increase with their dispersal rate.

Second, restoration cost declines little by introducing a species at a density below the (estimated) carrying capacity, unless the species disperses very slowly. Third, a uniform spatial distribution within the initial populations's cluster adds little or no proportional cost over the ogive profile assumed in our simulated annealing method, as long as the sex ratio is close to optimal. Spatial uniformity will likely prove more practical for most animals. Given a uniform density close to the positive, stable equilibrium, restoration should focus on an initial population whose expanse exceeds the critical-cluster size, which (again) increases with dispersal rate.

Finally, adjusting frequencies of the sexes in an initial population may decrease the cost of successful restoration. Of course, if one sex always limits population growth, an excess of that sex promotes restoration. If population growth depends on the density of each sex, introducing the sexes at different densities, or with different cluster sizes, may prove advantageous. Sex ratio at birth may be unbiased, but mortality rates may differ between sexes, particularly during dispersal. Adjusting the sex ratio at introduction to match frequencies at positive equilibrium densities should promote successful restoration, and reduce its cost.

\section*{Acknowledgments}

The National Science Foundation supported this research under Grant Nos. DEB-0918413 and DEB-0918392. We thank George Robinson and Zoltan Racz for discussion.

\section*{Author contributions}

Conceived and designed the experiments: FM TC GK.
Performed the experiments: FM CC.
Analyzed the data: FM CC TC GK.
Wrote the paper: FM CC TC GK.

\clearpage

\renewcommand{\thefigure}{S\arabic{figure}}\setcounter{figure}{0}
\renewcommand{\theequation}{S\arabic{equation}}\setcounter{equation}{0} 
\renewcommand{\thesection}{S\arabic{section}}\setcounter{section}{0}
\renewcommand{\thesubsection}{\thesection.\arabic{subsection}}

\setcounter{page}{1}

\begin{flushleft}
{\Large
\textbf{Supporting Information}
\\
\vspace*{0.5cm}
Restoration Ecology: Two-Sex Dynamics and Cost Minimization
}
\\
F. Moln\'{a}r Jr., C. Caragine, T. Caraco, G. Korniss
\\
\end{flushleft}

\section{Fixed points of the single-sex model}

Recall the equation for the single sex model:
\begin{equation}
\partial_{t} u = D \nabla^{2}u + \frac{1}{2}u^{2}(1-2u) - \mu u
\label{eq-apx-simplemodel}
\end{equation}
At equilibrium, the left hand side of the equation equals zero:
\begin{equation}
0 = D \nabla^{2}u + \frac{1}{2}u^{2}(1-2u) - \mu u
\label{eq-apx-simplefix}
\end{equation}
We have the following fixed points:
\begin{eqnarray}
u^{*}_{1} & = & 0, \\
u^{*}_{2,3} & = & \frac{1}{4} \mp \sqrt{\frac{1}{16}-\mu}. \;
\label{eq-apx-fix1}
\end{eqnarray}
Subscript numbering of the fixed points matches the numbering of stationary-solution fixed points in the paper. $u^{*}_{1}$ is the stable zero fixed point.  $u^{*}_{2}$ is unstable, $u^{*}_{3}$ is stable, corresponding to the Allee threshold and positive stationary densities, respectively.

Since the discriminant in equation (\ref{eq-apx-fix1}) must be non-negative, we have a necessary condition for the existence of fixed points $u^{*}_{2}$ and $u^{*}_{3}$:
\begin{equation}
\mu < \frac{1}{16}.
\end{equation}

\section{Fixed points of the two-sex model}

For the two-sex model, our previous paper [1] provides derivations of the fixed points; we repeat them for the reader's convenience.
\begin{eqnarray}
\partial_{t} f & = & D_{f} \nabla^{2}f + \theta \left(1-m-f\right) f m - \mu_{\rm f} f \nonumber \\
\partial_{t} m & = & D_{m} \nabla^{2}m + \left(1-\theta\right) \left(1-m-f\right) f m - \mu_{\rm m} m \;.
\label{eq-apx-model}
\end{eqnarray}
We have a trivial fixed point at zero densities:
\begin{eqnarray}
f^{*}_{1} & = & 0, \\
m^{*}_{1} & = & 0. \;
\label{sa_fp1}
\end{eqnarray}
To obtain the non-trivial fixed points, we first manipulate the two
stationary state equations, obtained from Eqs.~(\ref{eq-apx-model}),
to write a simple quadratic equation for the stationary total
density,
\begin{equation}
N(1-N)=\frac{\mu_{\rm f}}{\theta} + \frac{\mu_{\rm m}}{1-\theta} \;,
\label{eq:sa_tot_fp}
\end{equation}
yielding solutions
\begin{equation}
N^{\pm} = \frac{1 \pm \sqrt{\mathbf{D}}}{2} \label{N_tot_sa}
\end{equation}
with
\begin{equation}
\mathbf{D}(\mu_{\rm f},\mu_{\rm m},\theta) = 1-4\left(\frac{\mu_{\rm
f}}{\theta} + \frac{\mu_{\rm m}}{1-\theta} \right) \;.
\label{D_sa_appendix}
\end{equation}
Finally, for the non-trivial female and male densities at
equilibrium, we have
\begin{equation}
(f^{*},m^{*})_{2,3} = \left(\frac{\mu_{\rm m}}{1-\theta} \cdot
\frac{1}{1-N^{\pm}}\;  ,\;   \frac{\mu_{\rm f}}{\theta} \cdot
\frac{1}{1-N^{\pm}} \right) \;. \label{sa_fp23}
\end{equation}

For $\mathbf{D} \geq 0$ all three fixed points are real.  The trivial (zero density) solution [Eq.~(\ref{sa_fp1})] and the ``$+$" solution
[Eq.~(\ref{sa_fp23})] are locally stable, separated by an unstable
(saddle) fixed point, the ``$-$" solution in Eq.~(\ref{sa_fp23}). The stability of these fixed points can be easily analyzed by linearizing Eqs.~(\ref{eq-apx-model}). For $\mathbf{D} < 0$, however, only one biologically meaningful (real) fixed point exists, the zero-density
solution [Eq.~(\ref{sa_fp1})], and extinction is always stable.

The biological significance of the structure of the above solutions
is two-fold [1]. First, for $\mathbf{D} > 0$, the system
exhibits the Allee effect.  Unless the (initial) population density
is sufficiently high ($N(t=0)>N^{-}$), the population goes extinct [1,2].
Second, provided that $\sqrt{\mu_{\rm f}} + \sqrt{\mu_{\rm m}}<1/2$,
there is a finite interval $\theta_{\rm c1}(\mu_{\rm f},\mu_{\rm
m})<\theta<\theta_{\rm c2}(\mu_{\rm f},\mu_{\rm m})$, where
$\mathbf{D}(\mu_{\rm f},\mu_{\rm m},\theta) > 0$, i.e., such that the population can persist at equilibrium. These boundaries are given by:
\begin{equation}
\theta_{\rm c1,2}(\mu_{\rm f},\mu_{\rm m}) = \frac{(1+4\mu_{\rm
f}-4\mu_{\rm m})\pm \sqrt{(1+4\mu_{\rm f}-4\mu_{\rm m})^2 -
16\mu_{\rm f}}} {2} \;.
\label{CultureConstraint1}
\end{equation}
Between the two critical points, at
\begin{equation}
\theta^{*}=\frac{1}{1+\sqrt{\mu_{\rm m}/\mu_{\rm f}}} \;,
\end{equation}
total density exhibits a maximum at
\begin{equation}
N^{\max}=N^{+}(\theta^{*})=\frac{1+\sqrt{1-4(\sqrt{\mu_{\rm
f}}+\sqrt{\mu_{\rm m}})^2}}{2}
\end{equation}
where the female to male density ratio is
$f^{*}/m^{*}=\sqrt{\mu_{\rm m}/\mu_{\rm f}}$.

\begin{figure}[h!]
\begin{center}
\includegraphics[width=4in]{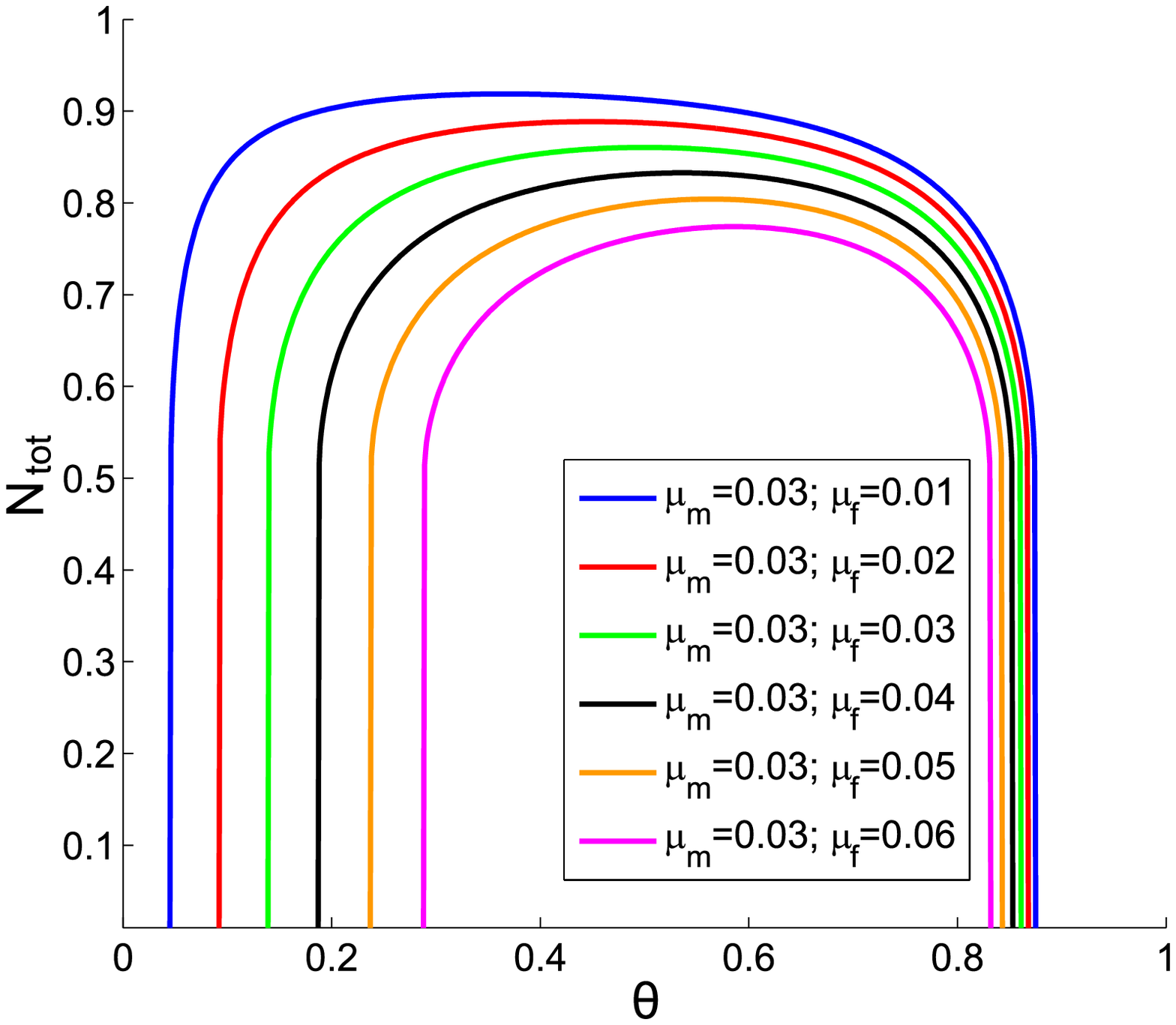}
\caption{Total population density \emph{vs} sex ratio, at various mortality rate combinations.}
\label{fig-ntot-theta}
\end{center}
\end{figure}

\section{Finding stationary solutions for single-sex model}

\begin{figure}[ht]
\begin{center}
\includegraphics[width=4in]{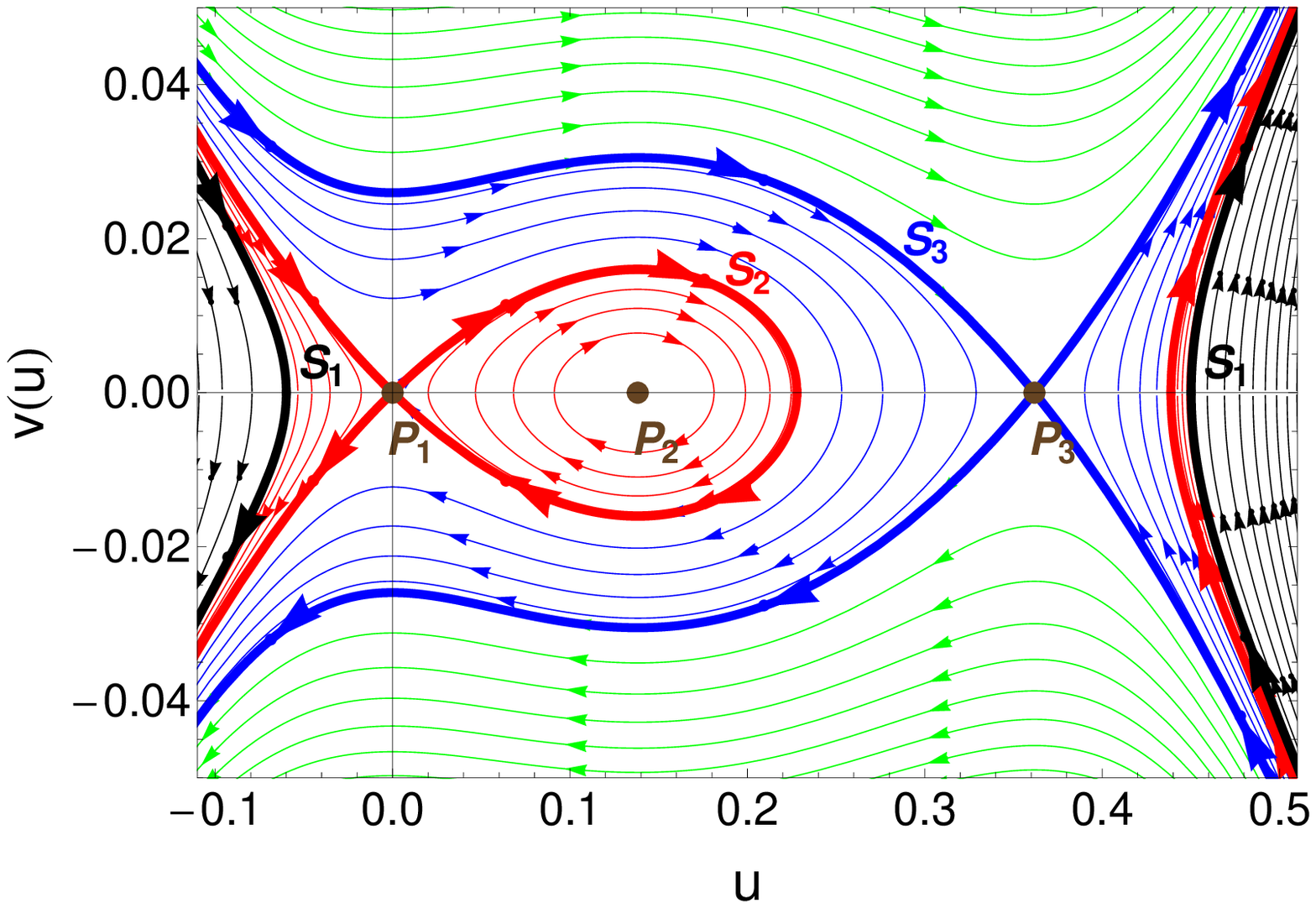}
\caption{Phase plot of dynamics in the single-sex model, described by Eq.~(\ref{eq-apx-vusol}). $D=1.0$, $\mu=0.05$. The dots indicate fixed points, the thick lines indicate separatrices. Different curves correspond to different $E$ parameters, however, values of $E$ were not chosen uniformly, for aesthetic reasons. This figure is also presented in the main text, it is repeated here for the reader's convenience.}
\label{fig-apx-uv}
\end{center}
\end{figure}

We need to solve the following equation to find stationary solutions (stable or unstable ones):
\begin{equation}
0 = D \nabla^{2}u + \frac{1}{2}u^{2}(1-2u) - \mu u.
\label{eq-apx-stac1}
\end{equation}
First we introduce $v$ as the first spatial derivative of $u$:
\begin{eqnarray}
\frac{\partial u}{\partial x} & = & v,  \\
\frac{\partial v}{\partial x} & = & \frac{-1}{D} \left( \frac{1}{2}u^{2}(1-2u) - \mu u \right). \;
\label{eq-apx-defv}
\end{eqnarray}
We can change variables to express $v(u)$ from $v(x)$, and rearrange the equation into this form:
\begin{equation}
v(u) \frac{\partial v}{\partial u} = \frac{-1}{D} \left( \frac{1}{2}u^{2}(1-2u) - \mu u \right).
\label{eq-apx-vu}
\end{equation}
By separating variables, we can find the following analytical solution:
\begin{equation}
v(u) = \pm \sqrt{ \frac{3u^{4} - 2u^{3} + 6 \mu u^{2} + 12 D E}{6D}},
\label{eq-apx-vusol}
\end{equation}
On the phase plane of this solution [Fig.~\ref{fig-apx-uv}] the fixed points correspond to homogeneous solutions, and curves correspond to inhomogeneous solutions. Specifically, closed curves represent periodic solutions.

We can find the special values for $E$ corresponding to separatrices, if we analyze local extremum points of $v(u)$. Moreover, to find such points, we need only study the fourth-order polynomial under the square root. Such a polynomial generally has three extrema. We utilize the phase plot [Fig.~\ref{fig-apx-uv}] to identify minima and maxima.

First, notice that when we are outside separatrix $S_{3}$ (green curves), $v(u)$ has a minimum at $P_{1}$, maximum at $P_{2}$ and minimum at $P_{3}$. The maximum at $P_{3}$ disappears when $S_{3}$ is reached. Therefore, we substitute the formula for $u^{*}_{3}$ (the fixed point at $P_{3}$) into the polynomial under the square root in $v(u)$ and equate to zero:
\begin{equation}
3(u^{*}_{3})^{4} - 2(u^{*}_{3})^{3} + 6 \mu (u^{*}_{3})^{2} + 12 D E = 0,
\label{eq-apx-sep3}
\end{equation}
and we solve for $E$. We denote the result with the subscript of the corresponding separatrix:
\begin{equation}
E_{3} = \frac{1}{24D} (u^{*}_{3})^{2} (u^{*}_{3} - 6 \mu)
\label{eq-apx-sep3sol}
\end{equation}
Therefore, if $E=E_{3}$, we obtain separatrix $S_{3}$.

Following the same logic, we find that the local minimum of $v(u)$ at $P_{1}$ becomes zero, when we reach separatrix $S_{2}$. We now substitute $u^{*}_{1}$ (the fixed point at $P_{1}$)  into the polynomial. Note that $u^{*}_{1} = 0$, which gives
\begin{equation}
12 D E = 0.
\label{eq-apx-sep2}
\end{equation}
Therefore, we obtain separatrix $S_{2}$ when $E=E_{2}=0$. Further, in this case we can find the zero point values of $v(u)$ easily:
\begin{equation}
3u^{4}-2u^{3}+6\mu u^{2} = 0.
\label{eq-apx-zeros}
\end{equation}
This gives the following zero points:
\begin{equation}
u_{1} = 0,
\label{eq-apx-zerosol1}
\end{equation}
\begin{equation}
u_{2,3} = \frac{1}{3} \pm \sqrt{\frac{1}{9}-2 \mu}.
\label{eq-apx-zerosol23}
\end{equation}
This gives the necessary condition $\mu < \frac{1}{18}$ for the existence of $S_{2}$ and the corresponding aperiodic solution. Note, that this is a stricter condition than the existence of nonzero fixed points, defined by equation (\ref{eq-apx-fix1}).

Separatrix $S_{1}$ is found again using the same method. It corresponds to the case when the local maximum at $P_{2}$ becomes zero (and disappears). We substitute $u^{*}_{2}$ into the polynomial to obtain:
\begin{equation}
3(u^{*}_{2})^{4} - 2(u^{*}_{2})^{3} + 6 \mu (u^{*}_{2})^{2} + 12 D E = 0
\label{eq-apx-sep1}
\end{equation}
We solve for $E$ and denote the result with the subscript number of the separatrix:
\begin{equation}
E_{1} = \frac{1}{48D} (u^{*}_{2} - 2 \mu) (u^{*}_{2} - 6 \mu).
\label{eq-apx-sep1sol}
\end{equation}
Therefore, if $E=E_{1}$ we obtain separatrix $S_{1}$. Note that $E_{1}$ is negative. To summarize, the separatrices have corresponding $E$ values that relate to each other as
\begin{equation}
E_{1} < E_{2}=0 < E_{3}.
\label{eq-apx-evalues}
\end{equation}

\section{Analyzing stationary solutions of the single-sex model}
Stationary solutions outsite $S_{2}$ separatrix (blue, green, and black curves on Fig.~\ref{fig-apx-uv}) have no physical or biological meaning, because they are unbounded, and extend to negative density ranges. Therefore we restrict our analysis to the periodic stationary solutions (red curves), and show that there is a minimum required spatial extent for their existence. First, we find the stability matrix and its eigenvalues.
\begin{eqnarray}
f(u,v) & = & \frac{\partial u}{\partial x} = v  \\
g(u,v) & = & \frac{\partial v}{\partial x} = \frac{-1}{D} \left( \frac{1}{2}u^{2}(1-2u) - \mu u \right)
\label{eq-jdef}
\end{eqnarray}
\begin{equation}
J = 
\begin{pmatrix}
\frac{\partial f}{\partial u} & \frac{\partial f}{\partial v} \\[0.3em]
\frac{\partial g}{\partial u} & \frac{\partial g}{\partial v} 
\end{pmatrix} 
 = 
\begin{pmatrix}
0 & 1 \\[0.3em]
\frac{3u^{2} - u + \mu}{D} & 0 
\end{pmatrix} 
\label{eq-jacobian}
\end{equation}
We obtain the following eigenvalues:
\begin{equation}
\lambda_{1,2} = \pm \sqrt{\frac{3u^2 - u + \mu}{D}}
\label{eq-eigenvalues}
\end{equation}
We are interested in the imaginary eigenvalues corresponding to the center ($P_{2}$):
\begin{equation}
\lambda_{P2} = \lambda_{1,2} \left( u_{2}^{*} \right) = \pm i \sqrt{\frac{16\mu -1 + \sqrt{1-16\mu}}{8D}}
\end{equation}
As we approach $P_{2}$, the orbits around it converge to a circle, with the angular frequency given by the imaginary eigenvalue:
\begin{equation}
\omega = \sqrt{\frac{16\mu -1 + \sqrt{1-16\mu}}{8D}}
\label{eq-freq}
\end{equation}
The periodic solution has the same angular frequency, therefore the wavelength in the limit of $P_{2}$ is:
\begin{equation}
L = \frac{2\pi}{\omega} = 2 \pi \sqrt{\frac{8D}{16\mu -1 + \sqrt{1-16\mu}}}.
\label{eq-length}
\end{equation}
This is the minimum habitat size allowing periodic stationary solutions. Note, that this case corresponds to $E = E_{1}$.

\section{Simulated annealing}
To move beyond the homogeneous distribution of the initial population as we seek to minimize cost, we need a method that finds the minimum cost in the domain of infinitely many possible distribution functions. The key to solving this problem is to define the initial distribution in discretized form. We can determine the eventual persistence or extinction of an initial population only by running the simulation to global equilibrium, and this simulation requires a spatial discretization for the integration. We use the same grid to discretize the initial distribution function. Further, we can assume that whatever initial distribution gives the minimum cost, it will have a finite support, therefore we can restrict the spatial extent of the distribution to a region in the center of the simulated space. If this restricted spatial domain includes $k$ grid points, we essentially reduce an infinitely detailed initial distribution to a $k$-dimensional function.

To find the minimum cost distribution, we use a global optimization method on the $k$-dimensional cost function, given the constraint that the population must persist.  We use simulated annealing [4], which has the ability to explore the multidimensional domain randomly without getting stuck in local minima. It slowly guides itself to the global minimum of the given function (in our case, the cost function) by lowering a control parameter, which slowly lowers the expected function values the random moves can take. Eventually, a global minimum is found.

Simulated annealing requires an initial guess for the minimum-cost shape, and we must define a transition function that can transform any shape to any other in some finite number of iterations.  At every step of the minimization, the new shape is proposed by applying the transition function to the previous shape. It is then accepted or rejected by the following criterion:
\begin{equation}
Pr(\mbox{accepted}) =
    \begin{cases}
        1 & \mbox{if } \Delta C < 0 \\
        \exp{ \left[ \frac{- \Delta C}{T} \right] } & \mbox{if } \Delta C \geq 0,
    \end{cases}
\label{eq-accept}
\end{equation}
where $\Delta C$ is the difference of cost between the newly proposed and previous functions, and $T$ is a temperature-like control parameter that is lowered over time by a given ``cooling schedule,'' so the probability of accepting changes that increase the cost function continuously declines. Eventually only changes that lower the cost are accepted, which finally moves the cost to a local minimum, but due to the gradual cooling and the stochastic nature of the procedure, this is also the global minimum with high probability.

Our minimization procedure starts with the following initial conditions:
\begin{eqnarray}
z(x) & := & \mbox{max}\left( 0, H \left( 1 - \frac{| (x-L/2) |}{W/2} \right) \right) \\
T(0) & := & 0.1
\label{eq-sa-initial}
\end{eqnarray}
Here, $T$ is the temperature parameter and $z(x)$ denotes either male or female initial density distributions. The shape of $z(x)$ is an isosceles triangle standing on its shorter side of length $W$ and having altitude $H$. These values are arbitrary, as long as the initial population they define generates a cost well above the sought minimum, so that they do not influence the minimization procedure. $L$ denotes the width of simulated habitat; $W < L$. In our simulations, we used the following values: $L=600$, $W=100$ (measured in discretized grid points), $H=0.4$, which is above the Allee threshold in every case we studied.

To minimize restoration cost with simulated annealing, we iterate the following steps. First, we determine the current spatial distribution's finite support:
\begin{eqnarray}
a & := & \{x : z(x) > \xi \land \forall i < x : z(i) \leq \xi \} \\
b & := & \{x : z(x) < \xi \land \forall i : a < i < b : z(i) > \xi \} \\
\widetilde{a} & := & a - (b-a)/4 \\
\widetilde{b} & := & b + (b-a)/4,
\label{eq-support}
\end{eqnarray}
where $\xi$ is a cutoff threshold set to $10^{-3}$. We use a randomized Gaussian function as transition function to generate a new proposed shape. The mean of this function is chosen from $[\widetilde{a}, \widetilde{b}]$ interval. Therefore, we allow for increasing the width of the density distributions, should the simulated annealing take that direction. The bell-shaped curve is defined by the following parameters:
\begin{eqnarray}
M & := & \widetilde{a} + \alpha (\widetilde{b}-\widetilde{a}) \\
A & := & 2 \beta T \\
V & := & (\widetilde{b}-\widetilde{a})/8 + 5 \gamma T (\widetilde{b}-\widetilde{a}) \\
S & := & \begin{cases}
        -1 & \mbox{if } \delta < 1/2 \\
        1  & \mbox{if } \delta \geq 1/2
        \end{cases} ,
\label{eq-transparams}
\end{eqnarray}
where $M$, $A$, $V$ and $S$ represent mean, amplitude, variance and sign, respectively, and $\alpha$, $\beta$, $\gamma$ and $\delta$ are uniform random numbers in the range $[0,1)$. Note, that amplitude and variance also depend on the current temperature, $T$, which helps the minimization process by making smaller changes as $T$ is reduced. The transition function is defined as
\begin{equation}
g(x) := \frac{S A}{\sqrt{2 \pi V}} \exp \left[ \frac{ -(x-M)^{2}}{2 V} \right].
\label{eq-transfunc}
\end{equation}
We then add $g(x)$ to the current shape, resulting in a new proposed shape:
\begin{equation}
\widetilde{z}(x) := \mbox{max}\left( 0, z(x)+g(x) \right).
\label{eq-proposed}
\end{equation}
We calculate shape distributions separately (independently) for both males and females. After the new shapes have been generated, the cost of the proposal is evaluated, and then accepted or rejected according to equation (\ref{eq-accept}). Finally, the temperature is lowered using a simple cooling schedule:
\begin{equation}
T(t+1) = 0.9999 T(t).
\label{eq-cooling}
\end{equation}
The iteration of these steps starts at $T(0)=0.1$ and continues until $T(t) < 10^{-4}$.

The constraint requiring successful restoration must be checked by running the simulation until convergence to a homogenous stationary state, persistence or extinction. If the newly proposed shapes result in extinction, the shapes are always rejected regardless of Eq.~(\ref{eq-accept}). Therefore, given that all previously accepted shapes resulted in survival, and given the convexity of the transition function (Gaussian), we know that when the newly proposed shapes increase the cost, then survival is guaranteed and there is no need to check it with a simulation. A test is only necessary when the cost is reduced. Still, this means that we need to run a numerical simulation for almost every second Monte Carlo step, which is computationally very intensive. To improve performance, we use GPGPU computation (using graphics processing units of video cards for general purpose computations) with CUDA [2]. Using GPUs can significantly improve the performance of PDE integration [3]. This technology lets us carry out each simulation within a fraction of a second, giving a total time for the simulated annealing in the order of a few hours.



\section*{References}
\begin{enumerate}

\item
Moln\'ar F Jr, Caraco T, Korniss G (2012)
Extraordinary sex ratios: cultural effects on ecological consequences.
PLoS ONE 7(8): e43364.

\item
NVIDIA Corp. (2013) CUDA C Programming Guide, \\
http://docs.nvidia.com/cuda/cuda-c-programming-guide/. Accessed: 01 June 2013.

\item
Moln\'{a}r F, Izs\'{a}k F, M\'{e}sz\'{a}ros R, Lagzi I (2011)
Simulation of reaction-diffusion processes in three dimensions using CUDA.
Chemometr. Intell. Lab. 108: 76-85.

\item
Kirkpatrick S, Gelatt CD, Vecchi MP (1983)
Optimization by Simulated Annealing.
Science 220: 671-680,

\end{enumerate}

\end{document}